# Asymmetric Oxidation of Giant Vesicles triggers Curvature-associated Shape Transition and Permeabilization.

**Heuvingh, Julien** [1]; **Bonneau, Stéphanie** [2]

1. Université Paris Diderot, PMMH, UMR7636 CNRS/ ESPCI/ Université Pierre et Marie Curie, 10 rue Vauquelin, Paris, France.
2. Université Pierre et Marie Curie, ANBioΦ, FRE3207 CNRS, 4 place Jussieu, Paris, France.

Both authors are corresponding authors
julien.heuvingh@espci.fr
stephanie.bonneau@upmc.fr

**Abstract (200 words)**

Oxidation of unsaturated lipids is a fundamental process involved in cell bioenergetics as well as in cell death. Using giant unilamellar vesicles and a chlorin photosensitizer, we asymmetrically oxidized the outer or inner monolayers of lipid membranes. We observed different shape transitions such as oblate to prolate and budding, which are typical of membrane curvature modifications. The asymmetry of the shape transitions is in accordance with a lowered effective spontaneous curvature of the leaflet being targeted. We interpret this effect as a decrease in the preferred area of the targeted leaflet compared to the other, due to the secondary products of oxidation (cleaved-lipids). Permeabilization of giant vesicles by light-induced oxidation is observed after a lag and is characterized in relation with the photosensitizer concentration. We interpret permeabilization as the opening of a pore above a critical membrane tension, resulting from the budding of vesicles. The evolution of photosensitized giant vesicle lysis tension was measured and yields an estimation of the effective spontaneous curvature at lysis. Additionally photo-oxidation was shown to be fusogenic.

**INTRODUCTION**

The oxidation of unsaturated lipids is of great interest, both from the biological and medical points of view (1-5). It may be generated in enzymatic or non-enzymatic reactions involving short-lived activated chemical species known as "reactive oxygen species" (ROS) (6). In cells, ROS are generated during the normal respiration process involving oxygen, oxidases and electron transport in mitochondria or the endoplasmic reticulum. Peroxidation of unsaturated lipids may be responsible, *in vivo*, for pathological processes such as drug-induced phototoxicity, arthrosclerosis and aging (3, 4). At cellular level, oxidation of lipids is involved in dysfunctions such as enhanced permeability, changes in membrane fluidity or release of lysosomal enzymes. In addition, the presence of oxidized phospholipids in lipidic membranes induces changes in their physical properties (7, 8).

The photochemical induction of oxidation is an effective way of inducing oxidation processes (6). It is supported by the ability of certain molecules, called photosensitizers, to generate ROS upon light irradiation. The specificity and preferential retention of certain photosensitizers by tumors, as compared to normal surrounding tissues, are the basis of an anti-tumoral therapy, the photodynamic therapy (PDT) (9). At high level, such light-induced molecular damage leads to the targeted cell's death. More recently, photosensitizer-induced lipid oxidation has been used to deliver macromolecular therapeutic agents to their intracellular targets by an approach called Photochemical Internalization (PCI) (10). After the uptake by endocytosis, the degradation of the macromolecules in lysosomes is greatly reduced by the photodynamic destabilization of the endoctytic vesicles membrane, increasing their biological activity.

The short half-life and limited diffusion length of the photo-induced ROS necessitate the close association of photosensitizers with the target site. For example, chlorin-generated subcellular singlet oxygen lifetime has been experimentaly measured between $4.5\pm0.5$ μs and $17\pm2$ μs (11). ROS typicaly diffuse less than 0.1 μm in a biological environment (12, 13). Due to the localized action of photosensitizers, the characteristics of their interaction with lipidic membranes are an important parameter controlling the effects of the photosensitizer-induced lipid oxidation (14). For tetrapyrrole photosensitizers, the ability to cross membranes is governed by the charge of their lateral chains (15-17). The chlorin e6 (Ce6), a second generation photosensitizer is not able to cross the biological membranes (18, 19). This allows an asymmetric labeling of the membrane, where the photosensitizer interacts only with the monolayer in contact with the photosensitizer solution. We thus labeled model membranes, Giant Unilamellar Vesicles (GUV) composed of dioleoylphosphatidylcholine (DOPC), which is an unsaturated lipid. Under light-induced oxidation, we observed morphological transitions and permeation of the GUV.

Morphological transitions in GUV have previously been observed in response to pH difference between compartments (20, 21), change in temperature (22), grafting of polymers on the membrane (23), ion adsorption on the lipid heads (24) or photoizomerisation of azobenzene-containing amphiphiles (25). One of the first models to account for various morphologies in vesicles was the spontaneous curvature model introduced by Helfrich (26). Increasing the spontaneous curvature induces membrane budding outside the GUV, whereas a decrease of the spontaneous curvature down to negative values induces a stomatocyte shape or budding inside the vesicle. Refinement of the theoretical comprehension of vesicle shape transitions led to the area-difference elasticity (ADE) model which takes into account the area difference between the two monolayers (27, 28). Any vesicle can then be defined by an area/volume ratio and a term of effective spontaneous curvature, which contains the area difference between the monolayers and the spontaneous local curvature. These two parameters give a phase diagram of vesicle shape that has been explored (22). For area/volume ratio corresponding to nearly spherical vesicles, an increase of the effective spontaneous curvature leads to shape transitions from an oblate ellipsoïd to a prolate ellipsoïd, and then to a pear and to a budding vesicle. Reducing the effective spontaneous curvature leads to a stomatocyte shape or inside budding.

DOPC membranes are essentially water permeable (40 µm/s (29)) and not permeable to polar solutes such as sugar at the time length involved in these experiments (6 $10^{-5}$ µm/s (30)). Due to osmotic equilibrium, the very slow permeability ensures a constant volume for vesicles. A much higher solute permeability is possible via pores, holes in the membrane allowing free flow. These pores have been observed in GUV (31, 32). They are energetically unfavorable because of the high cost of the exposition to water of hydrophobic lipid tails at the rim of the pore (33, 34).

In this work, we photochemically induce the lipid oxidation in DOPC-GUV. The photosensitizer used, Ce6, allows a symmetric as well as an asymmetric targeting of the membrane bilayers and a fine control of the location of the oxidation. The induced morphological transitions are corelated to the targeted leaflet and show a decrease in the spontaneous curvature of the targeted leaflet. The eventual permeabilization of the membrane has been measured and can be linked to the tension due to the budding of vesicles.

**MATERIALS AND METHODS**

*Chemicals.* All chemicals were purchased from Sigma (USA), except dioleoylphosphatidylcholine (DOPC) and dipalmitoylphosphatidylcholine (DPPC) from Avanti Polar Lipids (USA), and chlorin e6 from Porphyrin Products (USA). Chlorin stock solution (5 mM) was prepared in ethanol and keept at –18 °C. The experimental Ce6 aqueous solutions were prepared, used without delay and handled in the dark. The osmolarity of the solutions was checked with an osmometer (Roebling, Germany).

*GUV formation.* GUV were formed by the electroswelling method (35). DOPC in chlorophorm was deposed on ITO-covered glass plates. A chamber was made from two such glass plates and a Teflon spacer of 4 mm; the solvent was dried in vacuum. The chamber was filled then with a solution of 300 mM sucrose and an AC field of 1 Volt and 8Hz was applied between the plates for 4 hours. For DOPC versus DPPC comparison, electroformation took place at 50°C, above the transition temperature of DPPC.

For observation, the GUV were mixed with a 300 mM glucose solution. The density difference between sucrose and glucose caused the GUV to sediment to the bottom of the chamber. The difference in optical index between sucrose inside and glucose outside the vesicle allowed phase contrast microscopy observation.

*Chlorin labeling.* Giant Vesicles were asymmetrically labeled with chlorin. The chlorin molecules were present outside the vesicles, inside the vesicles or both outside and inside. The concentrations of chlorin used were 2.5, 5, 12.5, 25, 50 and 125 µM. The measured pH of the solutions was 5.1 ± 0.2. For chlorin present outside the vesicles, chlorin was diluted in the glucose solution before mixing with GUV. For chlorin present inside the vesicles, GUV were prepared with a sucrose solution containing chlorin in the chamber and subsequently rinsed. Rinsing was achieved by mixing the GUV with the glucose solution and carefully centrifuging them twice (25g for 20 min.). At the typical lipid concentration in GUV solution (~$10^{-6}$ M), the majority of the chlorin is unbound. This fact, combined with the high exchange rate of the chlorin with the bulk medium (19) ensures the exit of the sensitizer from the membrane outer leaflet and its rinsing. GUV with symmetrical chlorin distribution were prepared with a sucrose solution containing chlorin in the chamber and then mixed with a chlorin-containing glucose solution.

*Observation and illumination.* GUV were observed under a Nikon Eclipse TE2000-U inverted microscope equipped with a 1.3NA 100x oil objective. Illumination was provided by a 100 W Hg-arc lamp with a 465-495 nm bandpass filter. Images were acquired with a Basler A602f digital camera using Labview (National Instruments, USA). Images were analyzed using Image J (NIH, USA) and Scilab (INRIA, France, www.scilab.org).

*Tension measurement.* Micropipettes were made from borosilicate glass capillary GC100–15 tubing (1.0-mm outside diameter x 0.58-mm wall thickness x 7-cm length, Harvard apparatus Ltd., Kent, UK) using a pipette puller (Sutter instruments, model P-2000). A homemade microforge was used to tune their inner diameter from 2.3 to 5.3 µm. Pipettes were coated with bovine serum albumin and rinsed extensively. Prior to experiments, the pipettes were filled with the glucose solution. A suction pressure was applied in the pipette by hydrostatic pressure. The suction pressure P produces a tension τ in the membrane such as

$\tau = P \cdot D_P / 4(1 - D_P/D_V)$ (1)

where $D_P$ and $D_V$ are respectively the diameters of the pipette and of the vesicle (33).

*Statistics.* Correlation between the radius and permeabilization time of vesicle were analyzed by a t test on the regression analysis with a statistical significance of 0.01. When the two parameters correlated, the time was "corrected" i.e. deduced from the regression using a radius of 3.98 µm, which is the mean GUV radius of our samples. When no correlation was found, the average over the vesicles was simply used.

**RESULTS**

**Shape Transitions**
We prepared giant vesicles with different localizations of chlorin e6 regarding the leaflets. As chlorin molecules have a high affinity for lipids and do not flip-flop at the time scale involved in these experiments (18, 19), Ce6 labels only the leaflet directly in contact with the photosensitizer solution. Thus, GUV labeled with Ce6 in the outer, the inner and both leaflets were studied.

Upon illumination, all vesicles containing photosensitizers underwent major shape transitions within a few to a hundred seconds after illumination start (see Fig. 1). Vesicles with photosensitizer in the inner leaflet showed shape transitions from oblate to prolate ellipsoïds, pear shape deformations and budding of small vesicles outside the GUV. Vesicles in the prolate shape fluctuated during tens of seconds until the budding of vesicles put them out of prolate shape. Vesicles with photosensitizer in the outer leaflets typically showed deformation for a short time (< 2 seconds) followed by an invagination or budding of a small vesicle inside the GUV. The budded vesicles were visible as light dots contrasting with the GUV inner medium. Vesicles with Ce6 in both leaflets show all of the above-mentioned shape transitions. Finally, vesicles with no Ce6 showed none of these shape transitions for over 15 minutes. A quantification of these shape transitions is presented in Fig. 2. These results are in accordance with a lowered spontaneous curvature of the leaflet in which photo-oxidation occurs.

**Permeabilization**

Vesicles with chlorin in the outer or both leaflets undergo permeabilization. The contrast between the vesicle and the surrounding solution gradually fades away until there is no phase difference between inside and outside medium and the vesicle is barely observable by its membrane. This is a signature of the sucrose diffusion from the inner medium to the outer medium and its replacement by glucose from the outer medium (36). A small portion of the vesicles (<10%) burst or leaked abruptly in a non-gradual fashion. We quantified the contrast fading, which typically varies like a decreasing exponential (Fig. 3 left).

$$c = c_0 \exp(-(t-t_0)/\tau) \qquad (2)$$

where $c$ is the concentration of sucrose in the inner medium, $c_0$ its initial concentration, $t_0$ the initial time of the experiment and $\tau$ the characteristic time of the diffusion process.

In accordance with Eq. 2, we extracted a starting and a characteristic leaking time. The exponential fit is in good agreement with the data ($R_2 > 0.95$ for over 87% of the vesicles and $R_2 > 0.9$ for over 96% of the vesicles). The characteristic and starting times were compared with the vesicle radius. The starting time is statistically independent of the vesicle size, except at the highest concentration of Ce6 (50 and 125 µM outside the vesicle). The characteristic time depends on the vesicle radius at all concentrations. The dependence is stronger for high concentrations than for low ($R^2 = 0.69$ for 125 µM, $R^2 = 0.22$ for 2.5 µM).

The starting and characteristic permeabilization times decrease when the quantity of chlorin is increased for GUV with photosensitizers in the outer medium (Fig. 3 right). A plateau is evidenced over 25 µM. The permeabilization is slower for vesicles photosensitized on both leaflets. In the case of vesicles with photosensitizer only in the inner medium, vesicles are not permeabilized after 15 minutes, except for the higher concentration of 125 µM. At this concentration, permeabilization is 3 times longer for vesicles bearing Ce6 on their inner leaflet than for vesicles bearing Ce6 on their outer leaflet. All vesicles had a spherical shape (showing tension) before the start of permeabilization.

In order to verify that the permeabilization was mainly due to the oxidation of the lipid unsaturation, we compared the photosensitization effect on DOPC and DPPC, which is a saturated phospholipid. The experiment was conducted at 50°C, above the transition temperature of DPPC. The vesicles were in presence of 25 µM Ce6 and illuminated by a 100 W Hg-arc lamp during two seconds without any filter. Contrary to the DOPC-GUV, the DPPC-GUV did not permabilize for more than 5 minutes (see Fig. S1 in the Supporting Material). The DOPC-GUV lost half of their contrast in 44 s ± 25 s, whereas more than 70% of the DPPC-GUV retained more than half of their contrast 300s after illumination. No systematic shape changes were detected with DPPC-GUV.

**Lysis tension**

Micropipette experiments were conducted to estimate the lysis tension of photosensitized DOPC-GUV. Vesicles in presence of 50 μM Ce6 in the outer medium were held by a micropipette and a suction pressure between 50 and 500 Pascal was applied, corresponding to a membrane tension from 0.05 to 0.5 mN/m (see materials and methods section). We kept the vesicle's tension constant and illuminated the sample. After a few seconds of illumination the vesicle's integrity was compromised, i.e. it disappeared inside the pipette. This is interpreted as the opening of a hole in the membrane, allowing the inside solution to leave the vesicle and the membrane to be sucked inside the pipette. The time between illumination and permeabilization were recorded and is presented in Fig. 4. It can be thought as the evolution of lysis tension over time. The time at which the vesicles permeabilize in experiments without pipettes is plotted on the same graph. This time corresponds to the time at which vesicles tensed at 0.15 mN/m leaked. Vesicles tensed in presence of Ce6 without illumination did not break for several minutes.

**Fusion**

Fusion of giant vesicles in contact with each other occurred frequently for GUV of more than 10 μm radius at the highest concentrations (125 μM) in outer leaflets. The contacts between these GUV lead to fusion in 71% of the cases (35 fusion events). See Movie S1 in the Supporting Material. We observed less than 5% of fusion between vesicles of less than 10 μm radius at 50 μM Ce6 and below. A more precise quantification of fusion proved difficult as the number of contacts is highly dependant on the size and density of the vesicle preparation. After micropipette experiments, when the vesicles were gradually aggregated by the evaporation in the experimental chamber, fusion occurred at a dramatically accelerated rate when illuminated.

## DISCUSSION

**Photo-oxidation products**

The chlorin e6 is a photosensitizer: it interacts very efficiently with light to produce reactive species (singlet oxygen and radicals) from its long-life triplet state. In ethanol, its quantum yield for $^1O_2$ production is important (around 0.65 (37)). In unsaturated lipids, excitation of a photosensitizer generates peroxides (38), which is highly unstable in presence of any trace of transition metal (39). In our microscopy experiments, it will spontaneously decay to a free radical lipid, which pulls out a hydrogen from another unoxidized unsaturated lipid, creating a new free radical and a hydroxylated lipid. The combination of molecular oxygen with the new free radical leads again to the removing of a hydrogen from another lipid and produces again free radical and lipid peroxide. The lipid peroxidation is thus a radical chain reaction leading to the formation of intermediate hydroperoxide (Fig. 5). Hydroperoxyl group induces hydrophilicity in the chains of lipids.

If it remains in the hydrocarbon region of the membrane bilayer, hydroperoxyl group should drastically change the membrane architecture by increasing the cross-section area of its lipid tails. The peroxide lipid would therefore have a packing parameter above one and would increase the relaxed area of the leaflet in which they are present. However, indirect evidences on polyunsaturated lipids have suggested that the peroxide group may more likely be near the water/membrane interface (40). In this case, the peroxide lipid will have a packing parameter below one and will still increase the leaflet relaxed area. Additionally, in presence of trace amounts of catalytic transition metals (e.g. $Fe^{2+}$), the photo-oxidation of biological lipidic systems results in a myriad of secondary products (39, 41). For monounsaturated fatty acids

(like are the hydrocarbon chains of DOPC), these major final products correspond to the cleavage of the carbon chain near the initial position of the double bound and give an alkene or an aldehyde (41). Such cleaved-lipids present a strongly modified geometry as compared to DOPC. They correspond to a packing parameter below one and to a decrease of the relaxed area in the leaflet in which they are present. They are also known to favor pore formation (33).

**Changes in effective spontaneous curvature**

The light-induced shapes observed in the vesicles containing photosensitizer in the inner leaflet (prolate, pears, external budding) correspond to shapes of higher effective spontaneous curvature in the ADE model. The large and slow fluctuations observed in this case are also typical of an increase of the effective spontaneous curvature near the budding transition (42). The internal budding observed when the photosensitizer is in the outer leaflet corresponds to a lower effective spontaneous curvature. The variety of shape transitions observed is therefore qualitatively in accordance with a lowered effective spontaneous curvature of the leaflet which is photosensitized.

More precisely, the effective spontaneous curvature in the ADE model is composed of two terms which cannot be experimentally separated. The first is the spontaneous curvature of the lipids linked to their packing parameter; the second is the difference in area between the two leaflets. Increasing the area of the external leaflet or changing its lipids towards cone shapes will increase the effective spontaneous curvature, whereas decreasing the area of the external leaflet or changing its lipids towards inverted cone shapes will decrease the effective spontaneous curvature. The effect on the membrane effective spontaneous curvature will be inversed if the internal leaflet is changed.

The products of lipid peroxidation discussed above will have contradicting effects on the effective spontaneous curvature of the leaflet. Lipid peroxides, while having a negative packing spontaneous curvature, will raise the leaflet effective spontaneous curvature by increasing its area. Inversely, cleaved-lipids have a positive packing spontaneous curvature but will lower the leaflet effective spontaneous curvature by decreasing its area. We need to estimate the magnitude of these effects on GUV membranes.

The effective spontaneous curvature can be expressed in the ADE model as:

$$C_0^* = C_0 + \frac{\alpha\pi}{h}\frac{\Delta A_0}{A} \qquad (3)$$

where $\alpha$ is the ratio between the local and non-local bending rigidity, h is the thickness of a leaflet, $C_0^*$ and $C_0$ are respectively the effective spontaneous curvature of the membrane and the spontaneous curvature, and $\Delta A_0$ is the optimal area difference between the two monolayers and A is the area of the membrane (28).

To compare the effect of lipid geometry on leaflet area increase, we use a simple geometric model, in which the cross section area is increased from A/N to $(A + \Delta A_0)/N$ at the position of the lipid double bond (N is the number of lipids in a leaflet). The spontaneous curvature of this lipid will then be equal to

$$C_0 = \frac{1}{h}\frac{\Delta A_0}{A} \qquad (4)$$

and the effective spontaneous curvature to

$$C_0^* = \frac{(\alpha\pi - 1)}{h}\frac{\Delta A_0}{A} \qquad (5)$$

This curvature is positive if $\Delta A_0$ is positive for DOPC as $\alpha = 1.2$ (22). The area increase is therefore dominant over the lipids packing spontaneous curvature.

The formation of lipid peroxide consequently increases the leaflet effective spontaneous curvature. This is true whether the peroxide is located at the water/membrane interface or in the membrane bulk. Likewise, the formation of cleaved-lipids will decrease the leaflet effective spontaneous curvature. Although both of these chemical products will likely be present in the photosensitized membrane, our results indicate that the cleaved-lipids have the dominant effect on membrane spontaneous curvature, triggering the morphological transitions described here.

**Permeabilization**

Permeabilization of photosensitized GUV was observed via the contrast fading as sucrose left the vesicle interior. The sucrose flux through a circular opening in the membrane is:

$$\Phi_v = 2Dsc \quad (6)$$

where D is the diffusion coefficient of sucrose, s the radius of the aperture, and $c$ the concentration in the inner medium (see Eq. 2) (34, 38). We assumed the external sucrose concentration to be zero, as the internal volume is negligible compared to the outer volume. The concentration in the inner medium obeys $dc = \Phi_v \, dt / V$ (V is the volume of the vesicle) and decreases exponentially with a characteristic time of V/2Ds.

The experimental data from the permeabilizations of individual GUV are indeed well fitted by such a decreasing exponential as shown in Fig. 2. The fact that permeabilization occurs suddenly, after 20 to 100 seconds of exposition, and the following exponential decay in sucrose concentration are both in good agreement with a pore opening scenario. Based on the above model, the diameter of a pore is estimated to be between 16 nm and 43 nm for a typical vesicle (τ between 10.4 s and 27.5 s for a chlorin concentration above 12.5 µM; R = 4 µm). These pore sizes are in the same range as the ones measured in electroporation of red blood cell ghosts (43) or the ones calculated for stretched giant vesicles (44).

**Permeabilization and tension**

Permeabilization and formation of a pore can be triggered by membrane tension. Membrane tension can be described as the combination of two contributions: an entropic tension due to the damping of thermal fluctuation modes, and an elastic term coming from the increased distance between the lipids. When measuring the area variation of a GUV while increasing the membrane tension (via micropipettes), these two contributions are visible successively. First the membrane entropic "ruffles" are smoothed, corresponding to a soft exponential rise in area, then the membrane stretches, corresponding to a linear increase of area (45). Eventually the stretching will break the membrane above a critical stress. The lysis stress for DOPC-GUV was measured to be 9.9 ± 2.6 mN/m (29). Unsurprisingly, photo-oxidation lowers the critical stress of DOPC vesicles as shown in our experiments. We observe that the mean time of membrane rupture during photo-oxidation decreases when the tension is increased, which is interpreted as a reduction of lysis tension with the photo-oxidation duration. The starting time of vesicle permeabilization when no suction is externally applied corresponds to a lysis tension of 0.15 mN/m. Therefore we hypothesize that when no external tension is applied, the vesicle tenses itself up to a tension of 0.15 ± 0.05 mN/m at which it lyses. We will show in the next paragraph how this tension can be reached through shape transition and budding. The fact that it takes longer for a vesicle to lyse when an external tension below 1 mN/m is applied can be explained by a partial suppression of budding due to this preexisting tension (46). Interestingly, the measured lysis tension of 0.15 mN/m corresponds to the transition between thermal smoothing and stretching (see Fig. 2 in (45)), i.e. lysis occurs at the slightest stretching.

**Budding and Tension**

In our experiments, the lipid oxidation on a GUV's external leaflet triggers a shape transition from a prolate form to a closed stomatocyte or internal budding. The volume/area ratio of the vesicle after internal budding will be that of a sphere. The ADE model, describing shape transitions due to curvature modifications, assumes the area to be constant and does not take into account any stretching of the membrane. Such description is perfectly valid for moderate changes in the curvature. However above a certain level of curvature, the energy released from the budding will become important compared to the energy in the thermal fluctuations of the membrane. The vesicle would then be able to bud by tensing itself and taking area from the ripples of the membrane. We compare these two energies, in a manner similar to (47), in order to estimate the minimum effective spontaneous curvature required to tense the vesicle up to its lysis tension $\sigma$.

The energy gained when budding a vesicle of radius r from a vesicle of radius R (r<<R) is:
$$H_b = 8\pi k_c (1 - C_0^* r) \quad (7) \text{ (see appendix)}$$
where $k_c$ is the bending rigidity of the membrane (see detailed calculation in appendix).
The energy necessary to entropically stretch a membrane from a tension $\sigma_0$ to a tension $\sigma$ is:
$$H_S = A \frac{k_B T}{8\pi k_c} (\sigma - \sigma_0) \quad (8)$$
where A is the area of the giant vesicle, and $k_B$ is the Boltzmann constant.

According to our hypothesis, the area gain from thermal fluctuation will be equal to the area of the budding vesicle:
$$\Delta A_{fluct} = A \frac{k_B T}{8\pi k_c} \text{Ln}\left(\frac{\sigma}{\sigma_0}\right) = 4\pi r^2 \quad (9)$$

Thus
$$\sigma_0 = \sigma \exp\left(-\frac{8\pi k_c}{k_B T} \frac{r^2}{R^2}\right) \quad (10)$$

For small buds (r<<R $(8\pi k_c/k_B T)^{1/2}$ ~250 nm), the stretching energy simplifies to :
$$H_S = 4\pi r^2 \sigma. \quad (11)$$
The total energy is then given by:
$$H = H_b + H_S = 4\pi r^2 \sigma + 8\pi k_c (1 - C_0^* r) \quad (12)$$
It will be minimum for $r = k_c C_0^* / \sigma$ at which
$$H = 8\pi k_c (1 - C_0^{*2} k_c / 2\sigma) \quad (13)$$
Budding will therefore be favorable for $C_0^* > (2\sigma / k_c)^{1/2}$
Therefore, budding can be responsible for the vesicle's tension increase up to the observed lysis tension of 0.15 ± 0.05 mN/m, if the effective spontaneous curvature is above 42 μm$^{-1}$ (±19 μm$^{-1}$). Expressed in terms of area difference, this corresponds to a decrease of 2.3% (±1.1%) of one of the monolayer area as compared to the other.

It is tempting to explain the different permeabilization behavior between GUV photosensitized on their outer or inner leaflet by the difference between external and internal budding. Due to external budding, GUV photosensitized on their inner leaflet will loose both area and volume, while GUV photosensitized on their outer leaflet will loose area and gain volume due to internal budding. This will result in more stretching when the outer leaflet is targeted than when the inner leaflet is. However, taking into account the size difference between the budding and the original vesicle (typically 1/5$^{th}$ as in Fig. 1), the change in volume is small compared to the change in area.

Recent publications showed that the GUV electroformation method used in these experiment can lead to a small degree of lipid peroxidation (8). The presence of peroxidized lipids in our vesicles prior to photosensitization should have no effect on our results as long as it is symmetric regarding the leaflet localization.

**Conclusion**
The asymmetrical shape transitions observed in photosensitized GUV reveal changes in their membrane spontaneous curvature. These modifications are in accordance with the presence of cleaved-lipids by-products of oxidation. Permeabilization and a decrease of lysis tension were also characterized. We developed a model linking the budding due to the spontaneous curvature change to a tension of the membrane up to the lysis level where membranes are permeated. These findings might shed a new light on some membrane permeation phenomenon involved in biomedecine photodynamic approaches and in cell oxidative stress.

**Aknowledgments**
The authors thank S. Cribier for the use of her micropipette device and C. Gourier for her microforge device. R. Santus, P. Sens, N. Puff, N. Coq for useful discussions.

**APPENDIX**
We calculate in this appendix the energy gained from budding a vesicle.
The ADE model Hamiltonian is:

$$H = \frac{1}{2}k_c \int (C_1 + C_2 - C_0)^2 dA + \frac{\alpha \pi k_c}{2Ah^2}(\Delta A - \Delta A_0)^2 \qquad (14)$$

Were $C_1$ and $C_2$ are the local curvatures of the vesicle, $C_0$ the spontaneous curvature of the membrane, $\Delta A$ is the area difference between the outer and the inner leaflet, and $\Delta A_0$ is the prefered area difference between the leaflets. Oxidation will change lipid area from the outer leaflet from $A$ to $A+\Delta A_0$, and the spontaneous curvature of the outer membrane lipids from essentially 0 to $C_0$.

For a sphere of radius R:

$$H_1 = \frac{1}{2}k_c \left(\frac{2}{R} - C_0\right)^2 A + \frac{\alpha \pi k_c}{2Ah^2}(8\pi Rh - \Delta A_0)^2 \qquad (15)$$

For a sphere of radius R' plus an internal bud of radius r<<R (we keep the total area constant so that $R'^2+r^2=R^2$):

$$H_2 = \frac{1}{2}k_c \left(\frac{2}{R'} - C_0\right)^2 4\pi(R^2 - r^2) + \frac{1}{2}k_c \left(\frac{2}{r} - C_0\right)^2 4\pi r^2 + \frac{\alpha \pi k_c}{2Ah^2}(8\pi R'h - 8\pi rh - \Delta A_0)^2 \qquad (16)$$

Assuming $1/R \ll C_0$, the energy difference when budding is then:

$$\Delta H = H_2 - H_1 = \frac{1}{2}k_c 4\pi r^2 \left(\left(\frac{2}{r} - C_0\right)^2 - C_0^2\right) + \frac{\alpha \pi k_c}{2Ah^2}\left((8\pi R'h - 8\pi rh - \Delta A_0)^2 - (8\pi Rh - \Delta A_0)^2\right)$$
$$\qquad (17)$$

It can be simplified as follows:

$$\Delta H = \frac{1}{2}k_c 4\pi r^2 \left(\frac{4}{r^2} - \frac{4C_0}{r}\right) - \frac{\alpha \pi k_c}{2Ah^2}16\pi rh \Delta A_0 \qquad (18)$$

$$\Delta H = 8\pi k_c \left(1 - C_0 r - \frac{\alpha \pi}{h}r\frac{\Delta A_0}{A}\right) \qquad (19)$$

$$\Delta H = 8\pi k_c \left(1 - C_0^* r\right) \tag{20}$$

with $C_0^* = C_0 + \dfrac{\alpha \pi}{h} \dfrac{\Delta A_0}{A}$ the effective spontaneous curvature.

# Figure

## Figure 1

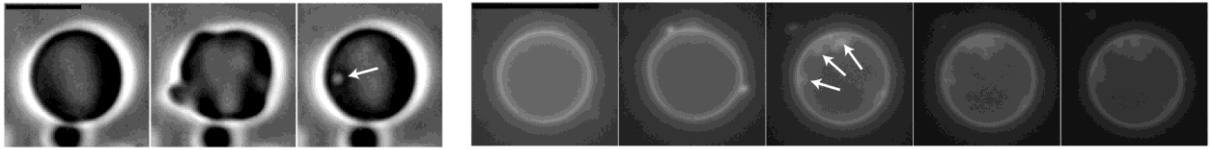

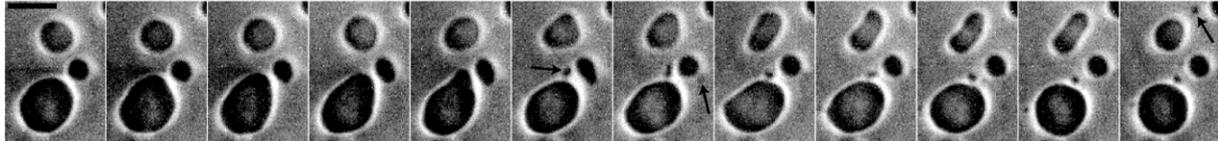

Time sequences of giant vesicles with photosensitizers targeted to the outer leaflet (upper row) or to the inner leaflet (lower row). The bar is 10 micrometers; the time lapse between two images is 1 second. In the first sequence of the upper row (phase contrast microscopy), a GUV photosensitized in its outer leaflet endures a short time deformation (second image) followed by a budding towards the inside (white arrow). In the second sequence of the upper row (fluorescence of chlorin) a GUV photosensitized in its outer leaflet endures a deformation (second image) followed by multiple budding towards the inside (white arrows). In the sequence of the lower row (phase contrast microscopy), three giant vesicles photosensitized in their inner leaflet show deformations: the upper vesicle exhibits a transition from oblate to prolate (due to the projection, oblate appears as circle and prolate as ellipse) whereas the two other vesicles are deformed into pear shapes. These shape transitions are followed by a budding to the outside on each GUV (black arrows). The morphology transitions observed when the outer leaflet is targeted denote a decrease of the membrane's equivalent curvature, while those observed when the inner leaflet is targeted denote an increase of the membrane's equivalent curvature.

## Figure 2

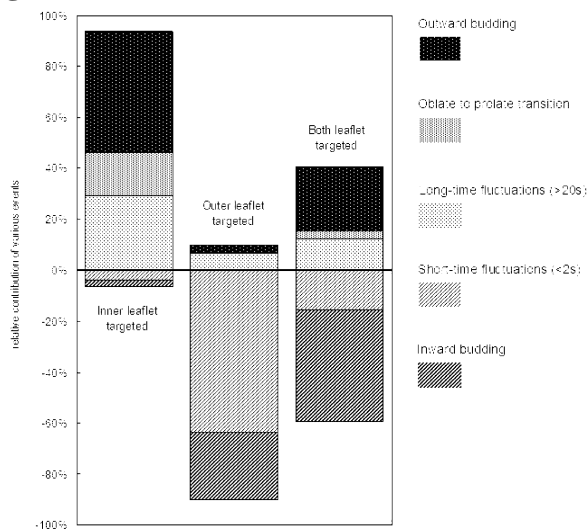

Proportion of each type of photoinduced morphology transition as a function of photosensitizer localization. From top to bottom, outward budding (white dots on black), oblate to prolate transition (close dots), long-time fluctuations (>20s) (spaced dots), short-time fluctuations (<2s) (wide hatching), inward budding (close hatching).

The mid-line between long-time and short-time fluctuations materializes the separation between events associated with an increase of the effective spontaneous curvature (above the line) and those associated with a decrease of the effective spontaneous curvature (below the line).

**Figure 3**

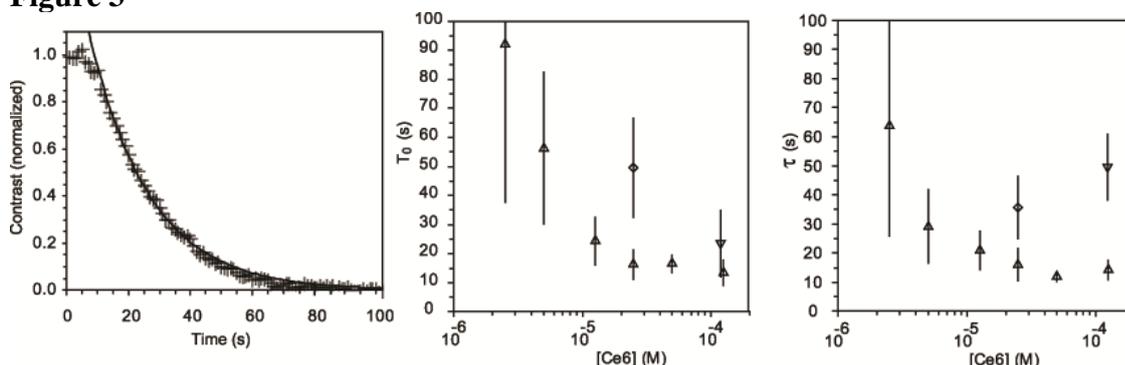

Permeabilization of vesicles. Left: typical evolution of the contrast between the inside and outside of a photosensitized vesicle at 25 µM Ce6 in the outer medium. The decay of the contrast shows equilibrium between the inside solute (sucrose) and the outside solute (glucose). The experimental points are fitted by a decreasing exponential, from which are extracted a starting and a characteristic times of permeabilization. Center and right: starting time ($t_0$) and characteristic time ($\tau$) of permeabilization for 50-120 vesicles for each concentration of Ce6. Upward triangles are vesicles with Ce6 in the outer medium, downward triangles are vesicles with Ce6 in the inner medium, and diamonds are in both media. The dependence on the vesicle radius is corrected (for $\tau$ at all concentrations and for $t_0$ at 50 and 125 µM outer medium)

**Figure 4**

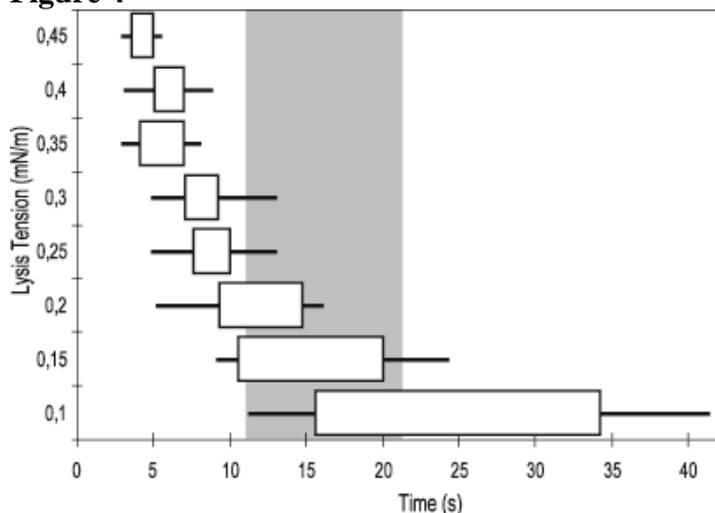

Evolution of lysis tension over time during photodamage. The time after the start of the illumination at which lysis occurs was recorded for different membrane tensions (113 vesicles total). The boxes boundaries are the upper and lower quartiles (median 50% inside the box). The whiskers boundaries are the upper and lower deciles (median 80% inside the whisker).

**Figure 5**

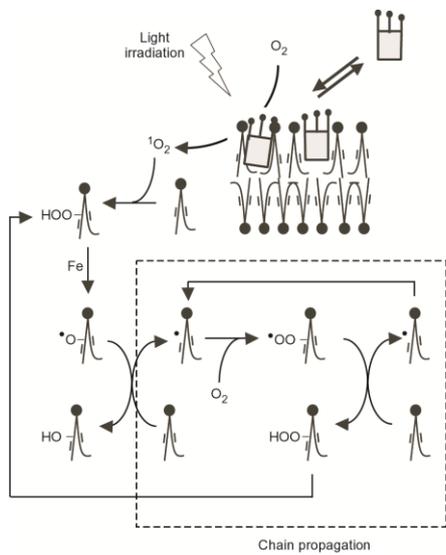
Diagram of the lipid photo-oxidation processes in the vesicle membrane. The lipid oxidation, initiated via singlet oxygen, is a chain reaction.